\documentclass[10pt,onecolumn]{IEEEtran}
\usepackage{amsmath,amsthm,amssymb}
\usepackage{enumerate}
\usepackage{color}
\usepackage{url}
\usepackage{balance}
\usepackage{graphicx} 
\usepackage{mathrsfs}
\usepackage{epsfig}
\usepackage{verbatim}
\usepackage{setspace}
\usepackage{stfloats}
\newtheorem{theorem}{Theorem}
\newtheorem{lemma}[theorem]{Lemma}

\newtheorem{claim}[theorem]{Claim}
\newtheorem{remark}[theorem]{Remark}
\renewcommand{\vec}[1]{\mathbf{#1}}

\newcommand{\cT}{\mathcal{T}}
\newcommand{\cP}{\mathcal{P}}
\newcommand{\cX}{\mathcal{X}}
\newcommand{\ctdX}{\wtd{\mathcal{X}}}
\newcommand{\cZ}{\mathcal{Z}}
\newcommand{\cU}{\mathcal{U}}

\newcommand{\cS}{\mathcal{S}}
\newcommand{\cJ}{\mathcal{J}}
\newcommand{\cY}{\mathcal{Y}}
\newcommand{\cB}{\mathcal{B}}
\newcommand{\cC}{\mathcal{C}}

\newcommand{\cL}{\mathcal{L}}

\newcommand{\bX}{\mathbf{X}}
\newcommand{\bZ}{\mathbf{Z}}
\newcommand{\bY}{\mathbf{Y}}
\newcommand{\bJ}{\mathbf{J}}
\newcommand{\bU}{\mathbf{U}}
\newcommand{\btdX}{\mathbf{\wtd{X}}}

\newcommand{\bx}{\mathbf{x}}
\newcommand{\bs}{\mathbf{s}}

\newcommand{\bT}{\mathbf{T}}
\newcommand{\bz}{\mathbf{z}}
\newcommand{\by}{\mathbf{y}}
\newcommand{\bj}{\mathbf{j}}
\newcommand{\bu}{\mathbf{u}}

\newcommand{\bbP}{\mathbb{P}}

\newcommand{\td}[1]{\tilde{#1}}
\newcommand{\wtd}[1]{\widetilde{#1}}

\newcommand{\bikd}[1]{\textcolor{blue}{#1}}
\newcommand{\removed}[1]{}

\newcommand{\nb}{(n)} 

\newcommand{\sT}{\mathscr{T}}

\newcommand{\sQ}{\mathscr{Q}}

\newcommand{\done}{\delta_1}
\newcommand{\dtwo}{\delta_2}
\newcommand{\dthree}{\delta_3}
\newcommand{\dfour}{\delta_4}

\newcommand{\dnot}{\delta_0}

\newcommand{\doned}{\delta_1(\delta)}
\newcommand{\dtwod}{\delta_2(\delta)}
\newcommand{\dthreed}{\delta_3(\delta)}
\newcommand{\dfourd}{\delta_4(\delta)}

\newcommand{\dnotd}{\delta_0(\delta)}

\newcommand{\fone}{f_1(\delta, \epsilon)}
\newcommand{\ftwo}{f_2(\delta, \epsilon)}

\newcommand{\tdftwo}{\td{f}_2(\delta, \epsilon)}
\newcommand{\gammad}{\gamma(\delta)}

\DeclareMathOperator*{\argmax}{arg\,max}

\title{Coding for Arbitrarily Varying Remote Sources}
\removed{
\author{
  \IEEEauthorblockN{Amitalok J. Budkuley\IEEEauthorrefmark{2} \thanks{\IEEEauthorrefmark{2}This work was done when Amitalok J. Budkuley was with the Dept. of Electrical Engineering at IIT Bombay, Mumbai-India.} }
	  \IEEEauthorblockA{
		The Chinese University of Hong Kong\\
		Shatin, Hong Kong\\
    Email: amitalok@ie.cuhk.edu.hk \vspace*{-0.5cm}	}
		\and
  \IEEEauthorblockN{Bikash Kumar Dey }
  \IEEEauthorblockA{
    Indian Institute of Technology Bombay\\
		Mumbai, India\\
    Email: bikash@ee.iitb.ac.in \vspace*{-0.5cm}}
	\and
  \IEEEauthorblockN{Vinod M. Prabhakaran}
  \IEEEauthorblockA{
     Tata Institute of Fundamental Research\\
     Mumbai, India\\
     Email: vinodmp@tifr.res.in \vspace*{-0.5cm}}
			
}
}

\author{
  \IEEEauthorblockN{Amitalok J. Budkuley\IEEEauthorrefmark{2} \thanks{\IEEEauthorrefmark{2}This work was done when Amitalok J. Budkuley was with the Dept. of Electrical Engineering at IIT Bombay, Mumbai-India.}, Bikash Kumar Dey and Vinod M. Prabhakaran}\\
  \IEEEauthorblockA{
    Emails: amitalok@ie.cuhk.edu.hk, bikash@ee.iitb.ac.in, vinodmp@tifr.res.in }
}

\begin{document}
\maketitle 
\thispagestyle{empty}
\interdisplaylinepenalty=0
\begin{abstract}
We study a lossy source coding problem for a memoryless remote source. The
source data is broadcast over an arbitrarily varying channel (AVC) controlled
by an adversary. One output of the AVC is received as input at
the encoder, and another output is received as side information 
at the decoder. The adversary is assumed to
know the source data non-causally, and can employ randomized jamming strategies
arbitrarily correlated to the source data. The decoder reconstructs
the source data from the encoded message and the side information.
We prove upper and lower bounds on the adversarial rate distortion 
function for the source under randomized coding. 
Furthermore, we present some interesting special cases of our general setup where the above bounds coincide, and thus, provide their complete rate distortion function characterization.
\end{abstract}
%
%
\section{Introduction}\label{sec:introduction}
\begin{figure}[!ht]
  \begin{center}
    \includegraphics[trim=0cm 11cm 0cm 0cm, scale=0.34]{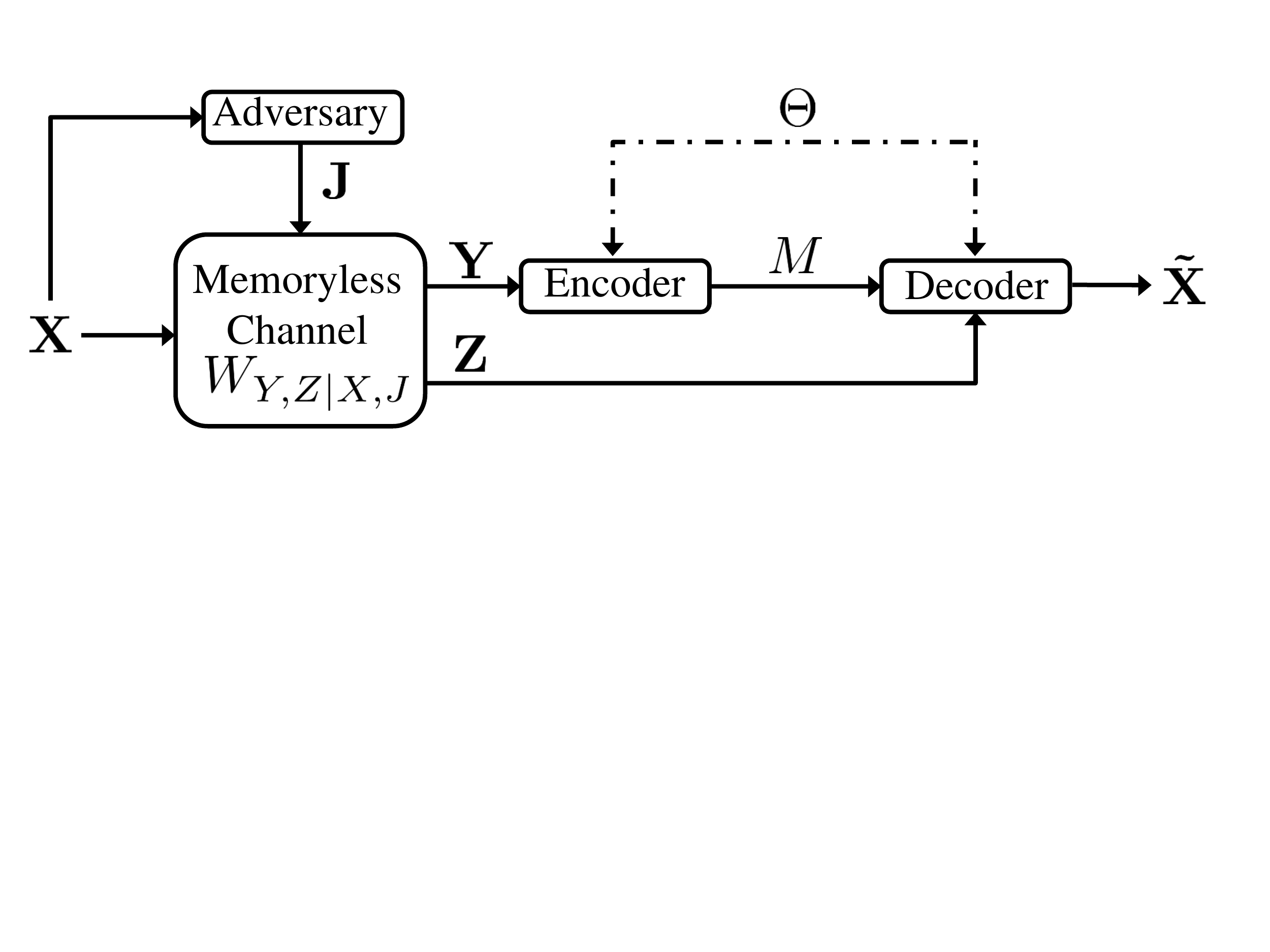}
    \caption{The arbitrarily varying remote source (AVRS) communication setup}
    \label{fig:main:setup:discrete}
  \end{center}
\end{figure}

Consider the communication scenario depicted in Fig.~\ref{fig:main:setup:discrete}. 
A memoryless source outputs an independent and identically distributed (i.i.d.) data vector $\bX$, which
is  broadcast over a  memoryless channel $W_{Y,Z|X,J}$.
Apart from the source, this channel has an input $\bJ$ from an adversary, 
and
it has two outputs $\bY$ and $\bZ$. The output $\bY$ is fed to the source
encoder which encodes it into a message $M$. The decoder receives the other
output $\bZ$ and the message $M$, and wants to reconstruct $\bX$ under
an average distortion criterion.
The adversary knows $\bX$ non-causally, and is allowed to employ randomized
vector jamming strategies arbitrarily correlated with it, thereby inducing an
arbitrarily varying channel (AVC)~\cite{lapidoth-narayan-it1998}. As is common
in AVC-related channel coding problems, we first undertake a study of this
setup under randomized coding in this paper, where we assume that the encoder-decoder share
an unbounded amount of randomness $\Theta$, unknown to the
adversary~\cite{lapidoth-narayan-it1998}. We
prove a maximin lower bound and a minimax upper bound for the rate distortion function for this \emph{arbitrarily varying remote source}  under randomized coding. 

In standard source coding scenarios involving noisy observations (e.g. noisy source
coding~\cite{dobrushin-tsybakov-it1962} or source coding with side information~\cite{wyner-ziv-it1976}), the noise statistics are known {\it a priori}. In our setup, however, the jamming signal of the malicious adversary renders these statistics completely arbitrary and  unknown, thereby making its analysis considerably more challenging. Furthermore, as depicted
in Fig.~\ref{fig:main:setup:discrete}, the jamming noise controls the observations $\bY$ as well as $\bZ$. Thus, the adversary in our problem can jointly degrade the compression as well as the
decoding/estimation phases of communication.

Lossy source coding has been studied extensively since the seminal work by
Shannon~\cite{shannon-ire1959}, and the field has subsequently been advanced in
many directions (cf.~\cite{kieffer-it1993,berger-gibson-it1998}). Apart from noisy source
coding~\cite{dobrushin-tsybakov-it1962} and source coding with side information~\cite{wyner-ziv-it1976}, some of
the other prominent directions related to this work include source coding under several distortion measures~\cite{dembo-weissman-it1993} and universal source coding~\cite{neuhoff-etal-it1975}. 
Particularly relevant are the \emph{compound} and \emph{universal} coding problem formulations 
 which have appeared for  classical coding, noisy/indirect coding, coding under 
several distortion measures, and coding with side information (cf.~\cite{sakrison-inc1969,dembo-weissman-it1993,watanabe-etal-it2014}, and
some of the references therein). Our problem also has a  direct connection to universal noisy source coding
problems which present a wider set of challenges
(cf.~\cite{dembo-weissman-it1993}) compared to their \emph{noise-free}
counterparts.
Another closely related model is that of an arbitrarily varying source (AVS)
introduced in~\cite{csiszar-korner-book2011}. This model is further studied 
under  variable rate codes in~\cite{feder-merhav-isit1995}. Inspired by an adversary capable of switching among different sources, Berger~\cite{berger-it1971} introduced a different AVS. In his problem, a multiplexer with inputs from several memoryless sources with
a common alphabet and a single output, feeds data to the encoder. The multiplexer is controlled by
a strictly causal switching adversary. An extension of results under  adversaries with causal as well as non-causal knowledge of the data has  subsequently appeared in~\cite{palaiyanur-it2011}. 

The rest of the paper is organized as follows. In
Section~\ref{sec:system:model}, we first introduce the notation and problem
setup. We state our main
result in Section~\ref{sec:main:results}. The proof of our main result is
presented in Section~\ref{sec:proof}. Finally, we discuss some implications of
our work, and make concluding remarks in Section~\ref{sec:conclusion}.
\section{Notation and Problem Setup}\label{sec:system:model}
\subsection{Notation and Preliminaries}
We denote random variables by upper case letters (e.g. $X$), the values they
take by lower case letters (e.g. $x$) and their alphabets by calligraphic
letters (e.g. $\mathcal{X}$). 
We use boldface notation to denote random vectors (e.g. $\vec{X}$) and their values (e.g. $\vec{x}$). Here, the vectors are of length $n$ (e.g. $\vec{X}=(X_1,X_2,\dots,X_n)$), where $n$ is the block length of operation. Let us also denote $\vec{X}^{i}=(X_1,X_2,\dots,X_i)$ and $\vec{x}^{i}=(x_1,x_2\dots,x_i)$ as well as  $\bX_{i}^{k}=(X_i,X_{i+1},\dots,X_k)$ and $\bx_{i}^{k}=(x_i,x_{i+1},\dots,x_k)$. We use the $l_{\infty}$ (denoted by $\|.\|_{\infty}$) norm for vectors.
For a set $\mathcal{X}$, let $\cP(\mathcal{X})$ be the set of all probability
distributions on $\mathcal{X}$. Similarly, let $\cP(\mathcal{X}|\mathcal{Y})$ be the set of all conditional distributions of a
random variable with alphabet $\mathcal{X}$ conditioned on another random
variable with alphabet $\mathcal{Y}$. For two random variables $X$ and $Y$,
we denote  the marginal distribution of $X$ obtained from the joint
distribution $P_{X,Y}$ by $[P_{X,Y}]_{X}$.  Distributions corresponding to
strategies adopted by the adversary are denoted by $Q$ instead of $P$ for
clarity. The set of all conditional distributions $\cP(\cJ|\cX)$ is
specifically denoted by $\sQ$. In cases where the subscripts are clear from the context, we sometimes
omit them to keep the notation simple. 
\removed{
Information quantities $H(X)$, $H(X,Y)$, $H(X|Y)$ and $I(X;Y)$ denote respectively the entropy of $X$, the joint entropy of $(X,Y)$, the conditional entropy of $X$ given $Y$ and the mutual information between $X$ and $Y$. 
}
Deterministic functions will be denoted in lowercase (e.g. $f$).  
We denote a
type of $X$ by $T_X$. Given sequences $\vec{x}$, $\vec{y}$, we
denote by $T_\vec{x}$ the type of $\vec{x}$, by $T_{\vec{x},\vec{y}}$ the joint
type of $(\vec{x},\vec{y})$ and by $T_{\vec{x}|\vec{y}}$ the conditional type
of $\vec{x}$ given $\vec{y}$. 
For $\epsilon\in(0,1)$, the set of $\epsilon$-typical set of $\vec{x}$ sequences for a distribution $P_X$ is
%
$\mathcal{T}^{n}_{\epsilon}(P_X)=\{\vec{x}:\|T_\vec{x}-P_X\|_{\infty}\leq \epsilon\}.$
%
In addition, for a joint distribution $P_{X,Y}$ and $\vec{x}\in\cX^n$, the conditionally typical set of $\vec{y}$ sequences, conditioned on $\vec{x}$, is defined as 
%
$\mathcal{T}^{n}_{\epsilon}(P_{X,Y}|\vec{x})=\{\vec{y}:\|T_{\vec{x},\vec{y}}-P_{X,Y}\|_{\infty}\leq \epsilon\}.$
%
%
\removed{
For any type $T_{X,Y}\in\mathscr{T}_{X,Y}^{\nb}$, we define
\begin{equation}\label{eq:ball}
B_{\delta}(T_{X,Y})=\{\tau\in\mathscr{T}_{X,Y}: \|\tau-T_{X,Y} \|_{\infty}\leq \delta\}.
\end{equation}
Note that if $T_{\vec{x},\vec{y}}\in B_{\delta}(T_{X,Y})$, then $(\vec{x},\vec{y})\in \mathcal{T}^n_{\delta}(T_{X,Y})$. Hence, for any $\tau\in B_{\delta}(T_{X,Y})$, we have
\begin{equation}\label{eq:size}
\big|  \{ \vec{x}: T_{\vec{x},\vec{y}}=\tau  \} \big| \leq 2^{n(H_{T_{X,Y}}(X|Y)+g(\delta))},
\end{equation}
where $g(\delta)>0$ and $g(\delta)\rightarrow 0$ as $\delta\rightarrow 0$.	
}
\subsection{The Problem Setup}
Refer the communication setup depicted in Fig.~\ref{fig:main:setup:discrete}.
Let $\mathcal{X}$, $\mathcal{Y}$, $\mathcal{Z}$, $\mathcal{J}$ and
$\mathcal{\wtd{X}}$ denote finite sets.  Consider an i.i.d. source with distribution $P_X$ and alphabet
$\cX$. We assume without loss of generality that $P_{X}(x)>0$, $\forall
x\in\mathcal{X}$. A length-$n$ block of data $\vec{X}$ is sent over a noisy
AVC. This channel has two inputs $X\in\cX$ and $J\in\cJ$ and two outputs
$Y\in\cY$ and $Z\in\cZ$, and its behaviour is given by the memoryless
distribution $W_{Y,Z|X,J}$. In Fig.~\ref{fig:main:setup:discrete}, the two inputs $\bX$ and $\bJ$ are from the source and
the jamming adversary respectively. 
The output $\bY$ is available at the encoder and $\bZ$ is available
at the decoder. 
We assume that the adversary knows $\bX$ non-causally.
%
Given inputs $\vec{x}$ and $\vec{j}$,  we observe $\vec{y}$ and $\bz$ with probability given by 
\begin{equation*}
\mathbb{P}\left( \vec{Y}=\vec{y},\vec{Z}=\vec{z}| \vec{X}=\vec{x},\bJ=\vec{j}\right)\hspace{-0.75mm}=\hspace{-0.75mm}\prod_{i=1}^n W_{Y,Z|X,J}(y_i,z_i|x_i,j_i).
\end{equation*}
%

The encoder compresses $\bY$ and transmits a message $M$ losslessly to the decoder. Using $M$ and the available side information $\bZ$, the decoder   outputs an estimate $\btdX$. The quality of the estimate is measured in terms of the average per-letter distortion 
%
$d(\vec{X},\vec{\wtd{X}})=\frac{1}{n} \sum_{i=1}^n d(X_i,\wtd{X}_i),$
%
where $d:\mathcal{X}\times \mathcal{\wtd{X}}\rightarrow  \mathbb{R}^+$ denotes  a single-letter distortion measure with $d_{\max}=\max_{(x,\tilde{x})\in \mathcal{X}\times\mathcal{\wtd{X}}} d(x,\tilde{x})<\infty$.

An $(n,R)$ \emph{deterministic code} of block length $n$ and rate $R$ is a pair
$(\psi,\phi)$ of mappings, consisting of the encoder map $\psi:\cY^n\rightarrow
\{1,2,\dots,2^{nR}\},$ and the decoder map $\phi: \{1,2,\dots, 2^{nR} \}\times
\cZ^n\rightarrow \ctdX^n.$ The encoder sends the message $M=\psi(\vec{Y})$ to
the decoder over an error free channel. An $(n,R)$ \emph{randomized code} of
block length $n$ and rate $R$ is a random variable which 
takes values in the set of $(n,R)$ deterministic codes. 
We denote by $(\Psi,\Phi)$ the encoder and decoder for this $(n,R)$
randomized code. This forms the shared randomness $\Theta$. The message 
sent is $M=\Psi(\vec{Y})$.
For this $(n,R)$ randomized code, the average distortion $D^{(n)}$ is given by 
\begin{equation*}
D^{(n)}=\max_{Q_{\vec{J}|\vec{X}}} \mathbb{E}[d(\vec{X},\Phi(\Psi(\vec{Y}),\vec{Z}))],
\end{equation*}
where the expectation is over the shared randomness $\Theta=(\Psi,\Phi)$, the source, the channel and the adversary's jamming action.
Given a target distortion $D$, a rate  $R$ is achievable if for any $\epsilon>0$ there exists an $n_0(\epsilon)$ such that for every $n\geq n_0(\epsilon)$ there exists an $(n,R)$ randomized code with the resulting average distortion $D^{(n)}\leq D+\epsilon$. We define the rate distortion function $R(D)$ as the infimum of all achievable rates.
Our aim is to determine the \emph{rate distortion function $R(D)$}.
\section{The Main Result}\label{sec:main:results}
%
%
Recall that $\sQ= \cP(\cJ|\cX)$  denotes the set of all conditional
distributions of $J$ given $X$.
For any distribution $Q_{J|X}\in\sQ$,
the system model gives the single-letter joint distribution
$P_XQ_{J|X}W_{Y,Z|X,J}$. 
Let
%
\begin{align}
D_0 &=  \min_{\tilde{x}(\cdot,\cdot)}\max_{Q_{J|X}\in \sQ} \mathbb{E}[d(X,\tilde{x}(Y,Z))]\label{eq:d0}\\
\text{and}\qquad D_1 &=  \min_{\tilde{x}(\cdot)}\max_{Q_{J|X}\in \sQ} \mathbb{E}[d(X,\tilde{x}(Z))].\label{eq:d1}
\end{align}
Here $D_0$ is the minimax average distortion
when both $\bY$ and $\bZ$ are available at the decoder, while $D_1$ is the minimax distortion when the decoder has access to only the side information $\bZ$ (Please see discussion in Sec.~\ref{sec:achieve}).
Let  $U$ be an auxiliary random variable with a 
finite alphabet $\mathcal{U}$ and conditional distribution $P_{U|Y}$,
such that $(X,J,Z) \leftrightarrow Y \leftrightarrow  U$ forms a Markov chain.
The joint distribution of $(X,J,Y,Z,U)$ is then given by
$P_XQ_{J|X}W_{Y,Z|X,J}P_{U|Y}$. 
We now define the following:
\begin{align}\label{eq:r:d:u}
R_U^*(D):=
\begin{cases}
 \displaystyle\min_{ P_{U|Y},\ \tilde{x}(\cdot,\cdot)  } \max_{Q_{J|X}\in \sQ  }I(U;Y|Z),\hspace{-2mm}& \mbox{if } D\in[D_0,D_1]\\
\hspace{5mm}0, &\mbox{if } D> D_1,
\end{cases}
\end{align}
where the minimization is over $P_{U|Y}\in\mathcal{P}(\mathcal{U}|\mathcal{Y})$ and 
$\tilde{x}:\mathcal{U}\times\mathcal{Z}\rightarrow \wtd{\mathcal{X}}$ such that
 $\mathbb{E}[d(X,\tilde{x}(U,Z))]\leq D,~\forall Q_{J|X}\in \sQ$. Clearly, we may restrict the cardinality of $U$ to $|\cU|\leq |\ctdX|^{|\cZ|}$ which is the number of possible functions from $\cZ$ to $\ctdX$.
\begin{align}\label{eq:r:d:l}
R_L^*(D):=
\begin{cases}
 \displaystyle\max_{Q_{J|X}\in \sQ  }\min_{P_{U|Y},\ \tilde{x}(\cdot,\cdot)} I(U;Y|Z), \hspace{-2mm}&\mbox{if } D\in[D_0,D_1] \\
\hspace{5mm}0, &  \mbox{if } D> D_1,
\end{cases}
\end{align}
%
where the minimization is over $P_{U|Y}\in\mathcal{P}(\mathcal{U}|\mathcal{Y})$
and $\tilde{x}:\mathcal{U}\times\mathcal{Z}\rightarrow \wtd{\mathcal{X}}$ such that $\mathbb{E}[d(X,\tilde{x}(U,Z))]\leq D$ for the specified $Q_{J|X}$. Here, we may restrict the cardinality of $U$ to $|\cU|\leq |\cY|+1$; this cardinality bound follows in a manner similar to~\cite{wyner-ziv-it1976}.
We next state our main result.
%
\begin{theorem}\label{thm:main:result:dmc}
The adversarial rate distortion function  $R(D)$ for the arbitrarily varying
remote source problem in Fig.~\ref{fig:main:setup:discrete} under randomized coding satisfies
\begin{equation*}
R_L^*(D)\leq R(D) \leq R_U^*(D).
\end{equation*}
\end{theorem}
Our setup can be considered to be an ``arbitrarily varying remote'' version 
of the Wyner-Ziv setup~\cite{wyner-ziv-it1976}, where both the input to the encoder 
 as well as the side-information are corrupted by the adversary. 
The setup gives two interesting special cases by limiting the adversary's
control to either one of these (i.e., $Y$ or $Z$).
If the adversary controls only $Y$, i.e., $W_{Y,Z|X,J}=W_{Y|X,J}P_{Z|X}$,
then the order of maximum and minimum can be interchanged. 
This is a consequence of the convexity-concavity properties of 
$I(U;Y) - I(U;Z)$. 
Specifically, $I(U;Y)-I(U;Z)$ is concave in $Q_{J|X}$ and
convex\footnote{In order to have a convex domain, we need to rewrite the
minimization as a minimization only over $P_{U|Y}$ where the alphabet of $U$ is
the set of Shannon strategies at the decoder; see~\cite{willems-report} for
details.} in $P_{U|Y}$. We can now use the Minimax
theorem~\cite{sion-pjm1958} to conclude that the minimax and the maximin are
equal. Similarly, if the adversary controls only $Z$, that is, when only the
side-information is arbitrarily varying ($W_{Y,Z|X,J}=P_{Y|X} W_{Z|X,J}$), then
again one can show that $I(U;Y) - I(U;Z)$ is convex and concave in $P_{U|Y}$
and $Q_{J|X}$ respectively. Hence, the maximum and minimum can be interchanged.
In both these special cases, the upper bound and the lower bound in
Theorem~\ref{thm:main:result:dmc} match, and they give a characterization of
the optimum rate.
\section{Proof of Theorem~\ref{thm:main:result:dmc}}\label{sec:proof}
\subsection{Achievability} \label{sec:achieve} We present an outline of the achievability proof. The detailed proof can be found in Appendix~\ref{app:ach:proof}.
%
%
Observe that if $D>D_1$, then we can estimate $\bX$ using
an estimator $\tilde{x}(z)$ based solely on the side information $\bZ$. 
Thus, for $D>D_1$, $R(D)=0$.

Let us now assume that $D_1\geq D \geq D_0$. 
We fix an arbitrary $P_{U|Y}$ and 
$\tilde{x}(u,z)$, and prove the achievability of the rate
\begin{align*}
R^{(P_{U|Y},\tilde{x})}:= &\max_{Q_{J|X}\in \sQ}I(U;Y|Z)\\
=& \max_{Q_{J|X}\in \sQ}(I(U;Y)-I(U;Z)),
\end{align*}
where the equality follows from the Markov chain $U\leftrightarrow Y\leftrightarrow Z$.
We rewrite this rate as~\footnote{Here we indicate $I(U;Y)$ as a function
of only $P_Y$ as $P_{U|Y}$ is fixed in our discussion of achievability. For the same reason,
we indicate $I(U;Z)$ only as a function of $Q_{J|X}$, as 
$P_X, P_{U|Y}$, and  $W_{Y,Z|X,J}$ are fixed in our discussion.}
\begin{align*}
&R^{(P_{U|Y},\tilde{x})}\\
& = \max_{Q_{J|X}\in \sQ} (I_{P_Y}(U;Y)-I_{Q_{J|X}}(U;Z))\\
& \geq \max_{Q_{J|X}\in \sQ} \max_{\begin{array}{c}P'_Y \in \cP(\cY) \\ P'_Y \stackrel{f(\epsilon)}{\approx} P_Y \end{array}}(I_{P'_Y}(U;Y)-I_{Q_{J|X}}(U;Z)) - \frac{\epsilon}{4},
\end{align*}
where we have $P_Y:=\left[P_XQ_{J|X}W_{Y,Z|X,J}\right]_Y$. Note that $P_Y$ is a
function of $Q_{J|X}$.
Here, we write $P_Y' \stackrel{f(\epsilon)}{\approx} P_Y$ to mean  that 
$||P_Y' - P_Y||_\infty \leq f(\epsilon)$. 
We have used a function $f(\cdot)$ such that $f(\epsilon)>0$ for $\epsilon >0$, and $|I_{P_Y'}(U;Y) - I_{P_Y}(U;Y)| \leq \epsilon/4$
if $P_Y' \stackrel{f(\epsilon)}{\approx} P_Y$. 
The existence of such a function follows from the uniform continuity
of $I(U;Y)$ as a function of $P_Y$ for fixed $P_{U|Y}$. Now interchanging the
maximizations, we get
\begin{align}
&R^{(P_{U|Y},\tilde{x})}\notag\\
& \geq \max_{P'_Y \in \cP(\cY	)} \max_{\begin{array}{c}Q_{J|X}\in \sQ \\ 
 P_Y \stackrel{f(\epsilon)}{\approx} P'_Y\end{array}}(I_{P'_Y}(U;Y)-I_{Q_{J|X}}(U;Z)) - \frac{\epsilon}{4} \notag\\
& \geq \max_{P'_Y \in \cP(\cY)}\hspace{-0.5mm} \left[I_{P'_Y}(U;Y)- \hspace{-1.5mm}\min_{\begin{array}{c}Q_{J|X}\in \sQ \\ P_Y \stackrel{f(\epsilon)}{\approx} P'_Y\end{array}}\hspace{-1.5mm} I_{Q_{J|X}}(U;Z)\right] - \frac{\epsilon}{4}. \label{eq:rds}
\end{align}
%
%
Now for every type $T_Y\in\cP(\cY)$, we define
\begin{IEEEeqnarray}{rCl}
R_U (T_Y) & :=& I_{T_Y}(U;Y)+ \frac{\epsilon}{4} \label{eq:ru}\\
\tilde{R}(T_Y)& :=& \min_{\begin{array}{c} Q_{J|X}\in \sQ\\ P_Y \stackrel{f(\epsilon)}{\approx} T_Y \end{array}} \hspace{-1.5mm}I_{Q_{J|X}}(U;Z) - \frac{\epsilon}{4}.
\label{eq:rtilde}
\end{IEEEeqnarray}%
\emph{Code Construction:}
\begin{itemize}
\item We will now describe the generation of a random code. We assume
that both the encoder and decoder share the ensemble of all possible such
codes, and they jointly select a code at random from this ensemble 
using their shared randomness $\Theta$. This is equivalent to generating the
code randomly and then sharing it between the encoder and the decoder.
\item For each type $T_Y\in\cP(\cY)$, we generate $2^{nR_U (T_Y)}$ vectors i.i.d. $\sim P_U$, where $P_U := \left[T_YP_{U|Y}\right]_U$,
to form the codebook $\cC(T_Y)$. The codebook $\cC(T_Y)$ is randomly partitioned into 
$2^{n(R_U (T_Y)-\tilde{R}(T_Y))}$ bins.
\item The randomly generated code containing the list of binned codebooks  
for each $T_Y$ is shared between the encoder and the decoder.
\end{itemize}
\emph{Encoder operations:}
\begin{itemize}
\item The encoder, upon observing a vector $\by$, computes its type $T_{\by}$.
It checks if there is at least one codeword in $\cC(T_{\by})$ which is jointly
typical with $\by$ with respect to (w.r.t.) the joint distribution $T_{\by}P_{U|Y}$.
The encoder then sends $T_{\by}$ and the bin index of such a
codeword in $\cC(T_{\by})$, selecting one uniformly at random if there is more than one possibility.
\item Since there are at most a polynomial number of types,
for large enough $n$, the rate required to convey $T_{\by}$ is at most
$\epsilon/4$. So, the rate of the full message is bounded as
\begin{align*}
R 
& \leq \max_{T_Y} (R_U (T_Y)-\tilde{R}(T_Y)) +\frac{\epsilon}{4}\\
& \leq R^{(P_{U|Y},\tilde{x})}+\epsilon. \hspace{10mm} \text{(using \eqref{eq:rds},~\eqref{eq:ru} and~\eqref{eq:rtilde})}
\end{align*}
\end{itemize}
\emph{Decoder operations:}
\begin{itemize}
\item The decoder knows $T_{\by}$ and the bin index sent by the encoder; it also knows $\bZ=\bz$ as the side information. 
The decoder identifies the set of conditional types
\begin{align*}
\sQ^{(n)}(T_\by) & = \{T_{J|X}\in \sQ : [P_X T_{J|X}W_{Y,Z|X,J}]_Y \hspace{0mm} \stackrel{f(\epsilon)}{\approx} T_\by\}
\end{align*}
such that the resulting $Y$-marginal distribution  is close to $T_{\by}$
\item 
The decoder then checks within the bin if there is a codeword $\bu$ such that $(\bu,\bz)$ is jointly typical w.r.t. the distribution 
$\left[P_XT_{J|X}W_{Y,Z|X,J}P_{U|Y}\right]_{U,Z}$ for some type $T_{J|X} \in \sQ^{(n)}(T_{\by})$. 
If there is a unique such codeword $\bu$, then it  chooses that codeword,
otherwise it chooses an arbitrary codeword $\bu$ 
from the bin. Using this codeword $\bu$ and $\bz$, it then outputs $\td{\bx}$, where $\td{x}_i=\td{x}(u_i,z_i)$, $i=1,2,\dots,n$.
\end{itemize}
\emph{Average distortion analysis:}
\begin{itemize}
\item We first analyse the error probability in decoding the right codeword
$\bu$. A decoding error can occur due to three possibilities:
\begin{enumerate}
\item Encoder does not find any codeword $\bu \in \cC(T_{\by})$ that is 
jointly typical with $\by$ w.r.t. $T_{\by}P_{U|Y}$. The probability 
that there is no such codeword in $\cC(T_{\by})$ is exponentially small (by covering lemma)
since $R_U(T_{\by}) = I_{T_{\by}P_{U|Y}}(U;Y) + \epsilon /4$.
\item Let us assume that the encoder succeeded in finding a suitable
codeword $\bu$. For this correct codeword $\bu$ and the actual 
conditional type $T_{\bj|\bx}$ instantiated
by the adversary, we will argue that $\bu$ will satisfy the decoding
condition with high probability (w.h.p.)~\footnote{All our w.h.p. statements
hold under ``except for an exponentially small probability.''}. 
First, w.h.p. $\by$ is typical w.r.t.
$[P_XT_{\bj|\bx}W_{Y,Z|X,J}]_Y$, i.e., $T_{\by}$ is ``close'' to
$[P_XT_{\bj|\bx}W_{Y,Z|X,J}]_Y$. In that case, 
$T_{\bj|\bx} \in \sQ^{(n)}(T_\by)$ is one of the conditional
types considered by the decoder for the code associated with $T_{\by}$.
Secondly, w.h.p. $(\by, \bu)$ is jointly typical w.r.t. $T_{\by}P_{U|Y}$ and so
it is also jointly typical w.r.t. the distribution $[P_XT_{\bj|\bx}W_{Y,Z|X,J}]_YP_{U|Y}$ (though
with a bigger slack). 
Now, using a version of the
refined Markov lemma~\cite[Lemma~5]{bdp-arxiv2016}, it follows that 
w.h.p., $(\bx,\bj,\by,\bu,\bz)$ is jointly typical w.r.t.
$P_XT_{\bj|\bx}W_{Y,Z|X,J}P_{U|Y}$.
In particular, $(\bu,\bz)$ is jointly typical
w.r.t. $[P_XT_{\bj|\bx}W_{Y,Z|X,J}P_{U|Y}]_{U,Z}$.
\item Now, let us consider all the wrong codewords in the bin.
For any type $Q_{J|X}\in \sQ^{(n)}(T_{\by})$, the probability that at least one of the wrong codewords will be jointly
typical with $\bz$ w.r.t. $[P_XQ_{J|X}W_{Y,Z|X,J}P_{U|Y}]_{U,Z}$ is exponentially
small due to the choice of $\tilde{R}(T_{\by})$ (by packing lemma). 
By taking union
bound over all (at most polynomial number of) types in $\sQ^{(n)}(T_{\by})$, the probability that any of them will be jointly
typical with $\bz$ w.r.t. $[P_XQ_{J|X}W_{Y,Z|X,J}P_{U|Y}]_{U,Z}$ for any such $Q_{J|X}$
is exponentially small.
\end{enumerate}
\item We now note that if $(\bx,\bj,\by,\bu,\bz)$ is jointly typical w.r.t. 
$P_XT_{\bj|\bx}W_{Y,Z|X,J}P_{U|Y}$, then $(\bx,\bj,\by,\bu,\bz,\td{\bx})$ is 
jointly typical w.r.t. 
$P_XT_{J|X}W_{Y,Z|X,J}P_{U|Y}\vec{1}_{\{\wtd{X}=\td{x}(U,Z)\}}$, and thus,
$(\bx, \td{\bx})$ is jointly typical.  
Finally, the average distortion $\mathbb{E}[d(\bX,\btdX)]$ is bounded
using the typical average lemma.
\end{itemize}

\begin{remark}
We have taken the code and binning rates (see~\eqref{eq:ru} and \eqref{eq:rtilde})
such that their difference is more than the $\max$ term in \eqref{eq:rds}.
A crucial feature of our achievability scheme is the choice of $\tilde{R}(T_Y)$
in \eqref{eq:rtilde}, which motivated the expression of 
$R^{(P_{U|Y}, \tilde{x})}$ as in \eqref{eq:rds}.
We now explain the insight behind this choice of $\tilde{R}(T_Y)$.
It is worth noting that instead of
taking the rate as the minimum  value of $I_{Q_{J|X}}(U;Z) - \epsilon/4$ over all $Q_{J|X}$
such that $[P_XQ_{J|X}W_{Y,Z|X,J}]_Y = T_Y$, we have taken it
to be the minimum over all $Q_{J|X}$ such that $[P_XQ_{J|X}W_{Y,Z|X,J}]_Y$ is 
``close'' to $T_Y$. Firstly, a part of our proof relies on bounding the probability
of error by union bounding over the (polynomial number of) 
conditional types (in $\sQ^{(n)}(T_\by)$)
that the decoder considers. For the union bound to work, the decoder
cannot consider every conditional  distribution $Q_{J|X}$,
which gives the right $P_Y$, as the number of such distributions can be
infinite.
Secondly, specially since the decoder only searches over the conditional
types and not every 
conditional distribution $Q_{J|X}$, it may not find any 
conditional type that gives exactly $T_\by$ as the marginal on $Y$.
Thirdly, our proof argument relies on the fact that 
the instantiated conditional type $Q_{\bj|\bx}$ 
is considered by the decoder, i.e., it is in $\sQ^{(n)}(T_\by)$.
However $[P_XQ_{\bj|\bx}W_{Y,Z|X,J}]_Y$ is only guaranteed (w.h.p.)
to be close to $T_\by$, and this is the reason for defining 
$\tilde{R}(T_Y)$ and $\sQ^{(n)}(T_\by)$ with a slack in the resulting
marginal on $Y$.
\end{remark}

\subsection{The proof of the lower bound} 
We will prove now that any achievable rate is lower bounded by
the maximin lower bound in~\eqref{eq:r:d:l}. We consider $D_1\geq D\geq D_0$. Consider an $(n,R)$ randomized code which achieves an average distortion of $D^{(n)}$, i.e., the code is such that
\begin{align*}
\frac{1}{n} \sum_{i=1}^n E[d(X_i,\widetilde{X}_i)] \leq D^{(n)}
\end{align*}
under any jamming distribution $Q_{\bJ|\bX}$. In particular, it satisfies the distortion constraint under the i.i.d. jamming distribution $Q_{J|X}$ with
\begin{align}\label{eq:mem:jam}
& Q_{\vec{J}|\vec{X}}(\vec{j}|\vec{x}) := \prod_{i=1}^n Q_{J|X}(j_i|x_i).
\end{align}
Under this jamming distribution, $(X_i,Y_i,Z_i)$, $\forall i$, form an i.i.d.
sequence with joint distribution given by $P_X P_{Y,Z|X}$, where
$P_{Y,Z|X}(y,z|x) = \sum_j W_{Y,Z|X,J}(y,z|x,j)Q_{J|X}(j|x)$, $\forall (x,y,z)$. Let us define
\begin{align}
F(D,Q_{J|X})&:=\min_{P_{U|Y},\ \wtd{x}(.,.)} I(U;Y|Z),
\label{eq:ef}
\end{align}
where the minimization is over $P_{U|Y}\in\mathcal{P}(\mathcal{U}|\mathcal{Y})$,
$\wtd{x}:\mathcal{U}\times\mathcal{Z}\rightarrow\wtd{\mathcal X}$
such that $\mathbb{E}[d(X,\tilde{x}(U,Z))]\leq D$ under the given $Q_{J|X}$.
It then follows using a similar argument as in the converse for the
Wyner-Ziv problem~\cite{wyner-ziv-it1976} that (see Appendix~\ref{app:sec:converse} for details)
\begin{align}
R \geq F(D^{(n)},Q_{J|X}).
\label{eq:wyn}
\end{align}
Hence, by the continuity of $F(D,Q_{J|X})$ in $D$ (Lemma~\ref{lem:f:d:props} in Appendix~\ref{app:sec:converse}),
\[ R(D) \geq F(D,Q_{J|X}).\]
Since this is true for any $Q_{J|X}\in\sQ$, we have the lower bound.


%
\section{Conclusion}\label{sec:conclusion} In this paper, we studied a setup of
lossy source coding for an arbitrarily varying remote source with
side-information.  As a natural first step, we gave upper and lower bounds for
the rate-distortion function for the randomized coding setup. The proof of
achievability employed novel techniques.  We also presented interesting special cases of our setup, and completely characterized their rate distortion function. The deterministic coding version is
open and is under current investigation.  
\section*{Acknowledgment}
This work was supported in part by Bharti Centre for Communication, IIT Bombay and in part by Information Technology Research Academy (ITRA), Government of India under grant ITRA/15(64)/Mobile/USEAADWN/01. In addition, Amitalok J. Budkuley, Bikash K. Dey and Vinod M. Prabhakaran were supported in part by RGC's GRF grants 14208315 and 14313116,  the Department of Science \& Technology, Government of India under a grant SB/S3/EECE/057/2013, and the Ramanujan Fellowship respectively. 
\appendices
\section{Proof of Achievability}\label{app:ach:proof}
In this detailed proof of achievability, we begin with the description of our randomized coding scheme.\\
\noindent \emph{Code Construction:}
\begin{itemize}
\item As discussed in the outline, the random code $\cC$ is a list of
individual codes $\cC(T_Y)$ for every type $T_{Y}\in\sT^{\nb}(\cY)$. This
list of codes is
shared as the common randomness $\Theta$ between the encoder-decoder.  
\item
For a fixed type $T_Y\in\sT^{\nb}(\cY)$, our code $\cC(T_Y)$ is a binned codebook
comprising $2^{n R_U(T_Y)}=2^{n(R(T_Y)+\tilde{R}(T_Y))}$ vectors
$\vec{U}_{j,k}$, where $j=1,2,\dots,2^{nR(T_Y)}$ and
$k=1,2,\dots,2^{n\tilde{R}(T_Y)}$. Here $R_U(T_Y)$ and $\tilde{R}(T_Y)$
are as given in \eqref{eq:ru} and \eqref{eq:rtilde} respectively, and
$R(T_Y) = R_U(T_Y) - \tilde{R}(T_Y)$. Every
codeword $\vec{U}_{j,k}$ is chosen i.i.d. $\sim P_U$, where $P_U:=[P_{U|Y}
T_{Y}]_{U}$. There are $2^{nR(T_Y)}$ bins indexed
by  $j$, with each bin containing $2^{n\tilde{R}(T_Y)}$ codewords indexed by
$k$. Let $\cB^{(T_Y)}_m$ denote the bin with index $m$.
Thus, our
code $\cC$ is the list containing $\cC(T_Y); T_Y\in\sT^{\nb}(\cY)$.
\end{itemize}
\emph{Encoding:}
\begin{itemize}
\item Given input $\bY$, the encoder determines its type $T_{\bY}$ to identify
$\cC(T_{\bY})$. In $\cC(T_{\bY})$, it finds a codeword  $\vec{U}_{m,l}$, where $m\in
\{1,2,\dots,2^{nR^{(T_{\bY})}}\}$ and $l\in
\{1,2,\dots,2^{n\tilde{R^{(T_{\bY})}}}\}$, such that 
\begin{equation}\label{eq:encoder:condition:dmc}
\|T_{\vec{U}_{m,l},\vec{Y}}-P_{U|Y}T_{\bY}\|_{\infty}\leq \delta_2(\delta).
\end{equation}
Here $\delta_2(\delta)>0$ is a fixed constant (the choice of 
$\delta_2(\delta)$ is indicated in Lemma~\ref{lem:covering})\footnote{Here $\delta>0$ is a function of $\epsilon$, such that
$\delta\rightarrow 0$ as $\epsilon \rightarrow 0$ and it is such that
\eqref{eq:D:eps} holds.}. This implies
that $\vec{U}_{m,l}$ and $\vec{Y}$ are jointly
typical according to the distribution $P_{U|Y} T_{\bY}$. If no such
$\vec{U}_{m,l}$ is found, then the encoder chooses $\bU_{1,1}$. If more than
one $\vec{U}_{m,l}$ satisfying~\eqref{eq:encoder:condition:dmc} exist, then the
encoder chooses one uniformly at random from amongst them. Let
$\vec{U}=\vec{U}_{M,L}$ denote the chosen codeword. 
\item The encoder transmits $T_{\by}$ and the bin index $M$ losslessly to the decoder. 
\end{itemize}
\emph{Decoding:}
\begin{itemize}
\item Let the bin index $m$ and side information $\vec{z}$ be received at the
decoder. In addition, the decoder knows the type $T_{\by}$ of
the encoder's input $\by$, and so the
code $\cC(T_{\by})$ used by the encoder. 
\item For some fixed parameter $\gamma(\delta)>0$
(the choice of $\gamma(\delta)$ is indicated in Lemma~\ref{lem:u:in:L}), the decoder determines the set of
codewords 
\begin{IEEEeqnarray}{rCl}\label{eq:dec:cond}
\mathcal{L}_{\gamma(\delta)}(m,\vec{z})=\Big\{ \vec{u}\in \mathcal{B}^{(T_{\by})}_m&: \|T_{\vec{u},\vec{z}}-[P_X T_{J|X} W_{Y,Z|X,J} P_{U|Y}]_{U,Z}\|_{\infty}\leq\gamma(\delta), \text{ for some } T_{J|X}\in \sQ(T_{\by}) \Big\}, 
\end{IEEEeqnarray}
%
Here $\sQ(T_{\by}):=\{T_{J|X}\in\sT^n(\cJ|\cX): [P_X T_{J|X} W_{Y,Z|X,J}]_Y\stackrel{f(\epsilon)}{\approx} T_{\by} \}$.  
%
%
%
\item If $\mathcal{L}_{\gamma(\delta)}(m,\vec{z})$ contains exactly one
codeword, then the decoder chooses it. Otherwise it chooses $\bu_{m,1}$. Let the
chosen codeword be $\vec{u}_{m,\tilde{l}}$. 
\item The decoder outputs $\td{\bx}$, where $\td{x}_i=\td{x}(u_i(m,\tilde{l}),z_i)$. 
\end{itemize}
\emph{Average distortion analysis:}

\removed{
\begin{align}
R_U (T_Y) & = I_{ P_{U|Y} T_Y}(U;Y)+ \epsilon/4 \label{eq:ru}\\
\tilde{R}(T_Y)& = \min_{ Q_{J|X}\in \sQ(T_Y)} I(U;Z) - \epsilon/4.
\label{eq:rtilde}
\end{align}
Note that since there are at most a polynomial number of types, 
for large enough $n$, the rate required to convey any $T_{Y}\in\sT^n(\cY)$ is at most
$\epsilon/4$. So, the rate $R$ is bounded as 
\begin{align}
R & \leq \max_{T_Y\in\sT^n(\cY)} (R_U (T_Y)-\tilde{R}(T_Y)) +\epsilon/4\notag\\
& \leq \max_{P_Y \in \cP(\cY)} \left(I_{P_Y}(U;Y)- \min_{Q_{J|X}\in \sQ(T_Y) }I_{Q_{J|X}}(U;Z)\right) + \epsilon/2 + \epsilon/4 \hspace{15mm}\text{  (using \eqref{eq:ru} and \eqref{eq:rtilde})}\notag \\
& \leq R^A(D) +\epsilon. \hspace{15mm} \text{(using \eqref{eq:rds})}   \label{eq:R:rate}
\end{align}
}
%
We first analyse the error in decoding the codeword $\bU=\bU_{M,L}$ chosen by the encoder. 
The decoder makes an error if one or more of the following events occur.
\begin{IEEEeqnarray*}{rCl}\label{eq:err:events}
E_{enc}&=&\{(\bU_{j,k},\bY)\not \in \cT^n_{\dtwo} (P_{U|Y} T_{\bY}), \forall j,k\}	\\
E_{dec_1}&=&\{(\bU,\bZ)\not \in \cL_{\gamma(\delta)}(M,\bZ) \}	\\
E_{dec_2}&=&\{(\bU_{M,l'},\bZ) \in \cL_{\gamma(\delta)}(M,\bZ) \text{ for some } l'\neq L\}	,
\end{IEEEeqnarray*}
Then, using the union bound we can express the probability of decoding error by
\begin{IEEEeqnarray}{rCl}\label{eq:P:E}
\bbP(E)\leq \bbP(E_{enc})+\bbP(E_{dec,1}|E^c_{enc})+\bbP(E_{dec,2}|E^c_{enc}).
\end{IEEEeqnarray}
We will show that for every $\epsilon>0$ there exists small enough $\delta>0$ 
such that $\bbP(E)\rightarrow 0$ as $n\rightarrow \infty$.
We first make the following obvious claim. 
\begin{claim}
Let $\bU$ be generated i.i.d. $\sim P_U$. Then, with probability at least $(1 - |\cU|e^{-2n\delta^2})$, $\bU \in 
\cT^n_{\delta}(P_U)$.
\end{claim}
Let us define this ``good'' event as $A_{U}:=\{\bU \in \cT^n_{\delta}(P_U)\}$.
We now state the following lemma which guarantees that the first term in~\eqref{eq:P:E} is vanishingly small.
\begin{lemma}\label{lem:covering}
Under the event $A_U$, there exist $\dtwod,\fone>0$, where $\dtwod,\fone\rightarrow 0$
as $\delta,\epsilon\rightarrow 0$,
such that the encoder
finds a codeword $\bU$ with probability at least $1-2^{-n\fone}$ such that
$(\bY,\bU)\in\cT^n_{\dtwo}(P_{U|Y}T_{\bY})$. 
\end{lemma}
The proof of this lemma follows from the covering lemma~\cite[Lemma~3.3]{elgamal-kim}. 
Note that this lemma specifies the $\dtwod$ parameter which appears in the definition of the encoder in~\eqref{eq:encoder:condition:dmc}.
This lemma implies $\bbP(E_{enc}) \rightarrow 0$ as $n\rightarrow 0$.
Our next lemma addresses the remaining two terms in the RHS of~\eqref{eq:P:E}.
\begin{lemma}\label{lem:u:in:L} Let the codeword chosen be $\vec{U}$ (where
$\vec{U}\in \mathcal{T}^n_{\delta}(P_U)$) and let the output on the channel
$W_{Y,Z|X,J}$ be $(\vec{Y},\bZ)$. Then, 

\begin{enumerate}[(a)]
\item there exists $\gamma(\delta)>0$, where $\gamma(\delta)\rightarrow 0$ as $\delta\rightarrow 0$, such that 
 except for an exponentially small probability, $\bU\in \cL_{\gamma(\delta)}(M,\bZ)$. 
\item there exists $f_2(\delta,\epsilon)>0$, where $f_2(\delta,\epsilon)\rightarrow 0$ as $\delta,\epsilon\rightarrow 0$, such that
\end{enumerate}
\begin{IEEEeqnarray}{rCl}
\mathbb{P}\left(\vec{U}_{M,l'}\in \mathcal{L}_{\gamma(\delta)}(M,\vec{Z}), \text{ for some } l'\neq L \right)\leq 2^{-n f_2(\delta,\epsilon)}.
\end{IEEEeqnarray}
\end{lemma}
The proof of this lemma can be found in Appendix~\ref{app:lem:u:in:L}. This
lemma specifies the parameter $\gamma(\delta)$ which appears in the 
the decoder operation in~\eqref{eq:dec:cond}. 
Lemma~\ref{lem:u:in:L} implies that $\bbP(E_{dec,1}|E^c_{enc}),\bbP(E_{dec,2}|E^c_{enc}) \rightarrow 0$ as $n\rightarrow 0$.
Hence, we can conclude that $\bbP(E)\rightarrow 0$ as $n\rightarrow
\infty$.

We now get a bound on the average distortion. Toward this, we first make the following claim.
\begin{claim}\label{claim:tdx}
There exists $r(\delta), f_3(\delta,\epsilon)>0$, where $r(\delta), f_3(\delta,\epsilon)\rightarrow$ as $\delta,\epsilon\rightarrow 0$, such that $\bbP\left((\bX,\btdX)\in \cT^n_{r(\delta)}(P_{X,\wtd{X}})\right)\geq 1-2^{-nf_3(\delta,\epsilon)}$.
\end{claim}
\begin{IEEEproof} 
By Claim~\ref{claim:x:2:u:typ} in App.~\ref{app:lem:u:in:L}, with high probability,
$(\bX,\bJ,\bY,\bZ,\bU)$ is $\delta_4$-typical according to the joint 
distribution $P_{X} T_{\bJ|\bX} W_{Y,Z|X,J} P_{U|Y}$.
As $\vec{\wtd{X}}$ is a deterministic function 
of $(\bU,\bZ)$, it follows by the conditional typicality lemma (see
Lemma~\ref{lem:cond:typ} in Appendix~\ref{app:lem:u:in:L}) that
with probability at least $(1-|\cX||\cJ||\cY||\cZ|
|\cU||\ctdX|2^{-n\delta_4^3})$, the tuple $(\bX,\bJ,\bY,\bZ,\bU,\btdX)$ is
$3\delta_4$-typical, and hence $(\bX,\btdX)$ is
$r(\delta)$-typical, where $r(\delta):=3|\cX| |\ctdX| \delta_4(\delta)$.
This completes the proof.
\end{IEEEproof}
We now show that the average distortion for the code $\cC$ can be made arbitrarily close to $D$. Let $\bar{E}:=\{(\bX,\btdX)\not \in \cT^n_{r(\delta)}(P_{X,\wtd{X}})\}$. From Claim~\ref{claim:tdx}, we know that $\bbP(\bar{E})\rightarrow 0$ as $n\rightarrow \infty$.
Then, 
\begin{IEEEeqnarray*}{rCl}
\mathbb{E}[d(\vec{X},\vec{\tilde{X}})]&=& \mathbb{P}(\bar{E})\mathbb{E}[d(\vec{X},\vec{\tilde{X}})|\bar{E}]+\mathbb{P}(\bar{E}^{c})\mathbb{E}[d(\vec{X},\vec{\tilde{X}})|\bar{E}^c]\\
&\leq& \mathbb{P}(\bar{E})\mathbb{E}[d(\vec{X},\vec{\tilde{X}})|\bar{E}]+\mathbb{E}[d(\vec{X},\vec{\tilde{X}})|\bar{E}^c].
\end{IEEEeqnarray*}
Recall that $d_{\max}<\infty$. In addition, from the typical average lemma 
we know that $\mathbb{E}[d(\vec{X},\vec{\tilde{X}})|\bar{E}^{c}]\leq D+h(\delta)$, where $h(\delta)>0$ and $h(\delta)\rightarrow 0$ as $\delta\rightarrow 0$. 
Thus, 
\begin{IEEEeqnarray*}{rCl}
\mathbb{E}[d(\vec{X},\vec{\tilde{X}})]&\leq& \mathbb{P}(\td{E}) d_{\max}+D+h(\delta)\\
&\stackrel{(a)}{\leq}& D+\epsilon. \yesnumber\label{eq:D:eps}
\end{IEEEeqnarray*}
As $\bbP(\bar{E})\rightarrow 0$ as $n\rightarrow \infty$, we choose a large enough $n$ and a small enough $\delta>0$ to get $(a)$. This implies that the average distortion can be made arbitrarily close to $D$. 
We have, thus, shown that for any $\epsilon>0$, the rate $R\leq\max_{Q_{J|X}}(I(U;Y)- I(U;Z))+\epsilon$ is achievable. This completes the proof of achievability.

\section{Proof of Lemma~\ref{lem:u:in:L}}\label{app:lem:u:in:L}
Let us define $\dnot=\delta/2$. Consider the  ``good'' encoder event
$E^c_{enc}=\{(\bY,\bU)\in\cT^n_{\dtwo}(P_{U|Y}T_{\bY})\}$. We now state and
prove some necessary claims.
\begin{claim}
Let $\bX$ be generated i.i.d. $\sim P_X$. Then, with probability at least $(1 - |\cX|e^{-2n\dnot^2})$, $\bX \in 
\cT^n_{\dnot}(P_X)$.
\end{claim}
Let us define this ``good'' event as $A_{X}:=\{\bX \in \cT^n_{\dnot}(P_X)\}$.
\begin{claim}\label{lem:j:typ:x}
Let $(\vec{x},\vec{j})$ be a pair of vectors where $\vec{x}\in\mathcal{T}_{\dnot}^n(P_X)$.
Then, $(\vec{x},\vec{j})\in \mathcal{T}_{\dnot}^n(P_{X} T_{\vec{j}|\vec{x}})$.
\end{claim}
Let us denote the event that $(\bX,\bJ)$ is jointly typical w.r.t. $P_X T_{\bJ|\bX}$ as $A_{X,J}$. By the above claim, we have $A_{X} \subseteq A_{X,J}$.
\begin{lemma}[Conditional typicality lemma]\label{lem:cond:typ}
Let $\bs\in\cT^n_{\dnot}(P_S)$ and $\bT$ be generated from $\bs$ using the memoryless distribution $W_{T|S}$. Then,
\begin{IEEEeqnarray}{rCl}
\mathbb{P}\left((\bs,\bT)\in\cT^n_{3\dnot}(P_S W_{T|X})\right)\geq 1-|\cS||\cT| e^{-2 n\delta^3_0}.\label{eq:P:s:t}
\end{IEEEeqnarray}
\end{lemma}
\begin{IEEEproof}
We need to show that
\begin{align*}
&\mathbb{P} \left(\left|T_{\bs,\bT}(s,t)-P_S(s) W_{T|S}(t|s) \right| > 2\dnot\right) 
\end{align*}
is exponentially small for all $s,t$.
We consider two cases. \\
{\it Case I:} $T_{\bs}(s)\leq \dnot$.
As $\bs\in\cT^n_{\dnot}(P_S)$, this implies that 
$ P_S(s)\leq T_{\bs}(s)+\dnot \leq 2\delta_0.  $
Then,~$\forall (s,t)$,
\begin{IEEEeqnarray*}{rCl}
\left|T_{\bs,\bT}(s,t)-P_S(s) W_{T|S}(t|s) \right| &=&\left|T_{\bs}(s)T_{\bT|\bs}(t|s)-P_S(s) W_{T|S}(t|s) \right|\\
&\leq& \max \left( T_{\bs}(s)T_{\bT|\bs}(t|s), P_S(s)W_{T|S}(t|s)\right)  \\
&\stackrel{}{\leq}& 2\dnot \cdot 1  \\
&=&2\dnot.
\end{IEEEeqnarray*}
Thus, for such $s$, $\bbP \left(\left|T_{\bs,\bT}(s,t)-P_S(s) W_{T|S}(t|s) \right| > 2\dnot\right)=0$.\\
{\it Case II:} $T_{\bs}(s)> \dnot$.
Using Chernoff-Hoeffding's  theorem~\cite[Theorem~1]{hoeffding-1963} for each $t\in\cT$, we have
\begin{IEEEeqnarray*}{rCl}
\mathbb{P}(|W_{T|S}(t|s)-T_{\bT|\bs}(t|s)| >\dnot, \text{ for any } t ) \leq |\cT|e^{-2n \delta^3_0 }.
\end{IEEEeqnarray*}
Now, it can be easily checked that 
$|W_{T|S}(t|s)-T_{\bT|\bs}(t|s)| \leq \dnot$ and $|P(s)-T_{\bs}(s)| \leq
\dnot$ together imply 
$$\left|T_{\bs}(s)T_{\bT|\bs}(t|s)-P_S(s) W_{T|S}(t|s) \right| \leq 2\dnot+\dnot^2 \leq 3\dnot.$$
Hence,~\eqref{eq:P:s:t} follows by taking union bound over all $s\in\cS$.
\end{IEEEproof}

\begin{claim}
With probability at least $(1-|\cX||\cJ||\cY||\cZ|e^{-2 \dnot^3 n})$,  $(\bX,\bJ,\bY,\bZ)$ are jointly $3\dnot$-typical according to the distribution $P_S T_{\bJ|\bX} W_{Y,Z|X,J}$.
\end{claim}
The proof of this result follows from Lemma~\ref{lem:cond:typ}.
We now consider this ``good'' event $A_{X,J,Y,Z}$, where $A_{X,J,Y,Z}:=\{(\bX,\bJ,\bY,\bZ)
\in \cT_{3\dnot}^n(P_X T_{\bJ|\bX} W_{Y,Z|X,J})\}$.
\begin{claim}
Under the event $A_{X,J,Y,Z}$, $\bY$ is $\done$-typical w.r.t. $P_Y=[P_XT_{\bJ|\bX}W_{Y,Z|X,J}]_Y$, where $\doned=3|\cX||\cJ||\cZ|\dnotd\rightarrow 0$ as $\delta\rightarrow 0$. That is,
$\|T_{\bY} - P_Y\|_{\infty} \leq 3|\cX||\cJ||\cZ|\dnot$. 
\end{claim}
The proof is straightforward, and hence, omitted. The above claim implies that, except for an exponentially small probability,
the decoder considers the conditional type $T_{\bJ|\bX}$ for decoding.
\begin{claim}\label{claim:y:u:tpy}
Under $E^c_{enc}$ and $A_{X,J,Y,Z}$, $(\bY,\bU)$ are jointly $\dthree$-typical according to the distribution 
$P_Y P_{U|Y}$, where 
$P_Y=[P_XT_{\bJ|\bX}W_{Y,Z|X,J}]_Y$ and
$\dthreed:=3|\cX||\cY||\cZ|\dnotd + \dtwod \rightarrow 0$ as $\delta \rightarrow 0$.
\end{claim}
\begin{IEEEproof}
Note that
\begin{align*}
\|P_YP_{U|Y} - T_{\bU\bY}\|_{\infty} & \leq \|P_YP_{U|Y} - T_{\bY}P_{U|Y}\|_{\infty}
               + \|T_{\bY}P_{U|Y} - T_{\bU\bY}\|_{\infty}\\
& \leq 3|\cX||\cY||\cZ|\dnot + \dtwo \quad \text{(using $E_{enc}$ and $E_{xjyz}$)}\\
& = \dthree,
\end{align*}
where $\dthree=3|\cX||\cY||\cZ|\dnot + \dtwo$.
\end{IEEEproof}
\begin{claim}\label{claim:unif:dist}
There exists $g(\delta)>0$, where $g(\delta)\rightarrow 0$ as $\delta\rightarrow 0$, such that $\forall \bu\in\cT^n_{\dthree}(P_{U|Y}P_Y|\by)$,
\begin{IEEEeqnarray}{rCl}\label{eq:tdX:dist}
P_{\bU}(\bU =\bu|\bY=\by)\leq 2^{-n(H(U|Y) -g(\delta) )},
\end{IEEEeqnarray}
where $H(U|Y)$ is computed with the distribution $P_{U|Y}P_Y$.
\end{claim}
\begin{IEEEproof}[Proof of claim]
We have two cases.\\
{\it Case 1:} When $\bu\in \cT^n_{\dtwo}(P_{U|Y}T_{\by}|\by) \bigcap \cT^n_{\dthree}(P_{U|Y}P_Y|\by)$.
Then we note that 
\begin{IEEEeqnarray*}{rCl}
&\mathbb{P} &\left( \bU =\bu|\bY=\by \right) \\
&\stackrel{}{=}& \mathbb{P} \left( \bU=\bu,\bU \in\cT^n_{\dthree}(P_{U|Y}T_{\by}|\vec{y})\big|\bY=\by  \right)\\
&=&\mathbb{P} \left( \bU \in\cT^n_{\dthree}(P_{U|Y}T_{\by}|\vec{y})\big|\bY=\by\right)\mathbb{P} \left( \bU=\bu\big|\bY=\by, \bU \in\cT^n_{\dthree}(P_{U|Y}T_{\by}|\vec{y}) \right)\\
&\leq& \mathbb{P} \left( \bU=\bu\big|\bY=\by, \bU \in\cT^n_{\dthree}(P_{U|Y}T_{\by}|\vec{y}) \right)\\
&=& \mathbb{P} \left( \bU_{1,1}=\bu\big|\bY=\by, \bU_{1,1} \in\cT^n_{\dthree}(P_{U|Y}T_{\by}|\vec{y}) \right)\\
&=& \frac{\mathbb{P} \left( \bU_{1,1}=\bu|\bY=\by\right)}{ \mathbb{P}\left(\bU_{1,1} \in\cT^n_{\dthree}(P_{U|Y}T_{\by}|\vec{y})|\bY=\by \right)}\\
&\stackrel{}{\leq}& 2\cdot \mathbb{P} \left( \bU_{1,1}=\bu|\bY=\by\right)
\quad \text{(since  $\mathbb{P}\left(\bU_{1,1} \in\cT^n_{\dthree}(P_{U|Y}T_{\by}|\vec{y})|\bY=\by \right) \rightarrow 1$ as $n\rightarrow \infty$)}\\
&\stackrel{}{\leq}& 2^{-n(H_{P_{U|Y}T_{\by}}(U|Y) -g_1(\dthree) )}
\end{IEEEeqnarray*}
where $g_1(\dthree)\rightarrow 0$ as $\dthree\rightarrow 0$.
Since $||P_Y - T_{\by}||_1 \leq |\cY|\cdot ||P_Y - T_{\by}||_\infty
\leq 3|\cX||\cJ||\cZ||\cY|\dnot$, and
$||P_YP_{U|Y} - T_{\by}P_{U|Y}||_1 \leq |\cU||\cY|\dthree$,
using \cite[Lemma~2.7]{csiszar-korner-book2011}, we get
\begin{IEEEeqnarray*}{rCl}
& |H_{P_Y}(Y) - H_{T_{\by}}(Y)| \leq 3|\cX||\cJ||\cZ||\cY|\dnot\cdot\log\left(
\frac{1}{3|\cX||\cJ||\cZ|\dnot}\right)\\  
& |H_{P_YP_{U|Y}}(U,Y) - H_{T_{\by}P_{U|Y}}(U,Y)| \leq |\cU||\cY|\dthree\cdot
\log\left( \frac{1}{\dthree}\right)  
\end{IEEEeqnarray*}
Together, the above two equations imply
\begin{align}
& |H_{P_YP_{U|Y}}(U|Y) - H_{T_{\by}P_{U|Y}}(U|Y)| \leq 2|\cU||\cY|\dthree\cdot
\log\left( \frac{1}{\dthree}\right). \label{eq:H:u:y}
\end{align}
By defining $g_2(\dthree) := g_1(\dthree) + 2|\cU||\cY|\dthree\cdot
\log\left( \frac{1}{\dthree}\right)$, we get
\begin{align*}
\mathbb{P} \left( \bU =\bu|\bY=\by \right)
& \stackrel{}{\leq} \frac{1}{2}\cdot 2^{-n(H_{P_{U|Y}P_{Y}}(U|Y) -g_2(\dthree) )}.
\end{align*}
{\it Case II:} 
When $\bu\not\in \cT^n_{\dtwo}(P_{U|Y}T_{\by}|\by) $. For such a $\bu$,
the encoder outputs it only if $\bU_{1,1}= \bu$ and there is no codeword
which is jointly typical with $\by$ w.r.t. $P_{U|Y}T_{\by}$.
Thus,
\begin{align*}
\bbP\left( \bU =\bu|\bY=\by \right)
& \stackrel{}{\leq} \bbP\left( \bU_{1,1} =\bu|\bY=\by \right)\\
& \leq 2^{-n(H_{[P_{U|Y}T_{\by}]_U}(U) - g_3(\dtwo))}\\
& \leq 2^{-n(H_{[P_{U|Y}P_Y]_U}(U) - g_4(\dtwo))},
\end{align*}
where $g_4(\dtwo) = g_3(\dtwo) + |\cU|^2|\cY|\dthree\cdot \log\left(
\frac{1}{|\cU||\cY|\dthree}\right)$.

Combining the two cases, and taking $g(\delta) = \max (g_2(\dthree),
g_4(\dthree))$, the lemma follows.
\end{IEEEproof}
\begin{lemma}[Refined Markov Lemma~\cite{bdp-arxiv2016}~\footnote{In the refined
Markov lemma presented in~\cite{bdp-arxiv2016}, condition (b) also has a lower
bound on $P_{\bZ}(\bz)$. However, the lower bound is not used in the proof
given in \cite{bdp-arxiv2016},
and hence, can be removed. Here, we state this lemma without any lower
bound. 
We also note that condition (a) and the upper
bound on the probability of a typical sequence imply that probability
of too many typical sequences can not be too small; and so some essence
of the lower bound in condition (b) is already implied by these. 
Thus, it is not surprising that the lower bound is not needed for the 
lemma to hold.
}]\label{lem:ref:markov:lemma}
Suppose $X\rightarrow Y\rightarrow Z$ is a Markov chain, i.e., $P_{X,Y,Z}=P_{Y}P_{X|Y} P_{Z|Y} $. Let $(\vec{x},\vec{y})\in \mathcal{T}^n_{\dnot}\left(P_{X,Y}\right)$ and $\vec{Z} \sim P_{\vec{Z}}$ be such that
\begin{enumerate}[(a)]
\item $\mathbb{P}\left((\vec{y},\vec{Z})\not\in
\mathcal{T}^n_{\dnot}\left(P_{Y,Z}\right)\right)\leq \epsilon$, where $\epsilon>0$,

\item for every $\vec{z}\in \mathcal{T}^n_{\dnot}\left(P_{Y,Z}|\vec{y}\right)$,
\begin{equation*}\label{eq:condition:markov}
P_{\vec{Z}}(\vec{z})\leq 2^{-n(H(Z|Y)-g(\dnot))},
\end{equation*}
for some $g:\mathbb{R}^{+}\rightarrow\mathbb{R}^{+}$, where $g(\dnot)\rightarrow 0$ as $\delta_0\rightarrow 0$.
\end{enumerate}
Then, there exists $\delta:\mathbb{R}^{+}\rightarrow\mathbb{R}^{+}$, where $\delta (\dnot) 
\rightarrow 0$ as $\dnot \rightarrow 0$, such that
\begin{equation*}
\mathbb{P}\left((\vec{x},\vec{y},\vec{Z})\not \in \mathcal{T}^n_{\delta(\dnot)}\left(P_{X,Y,Z}\right)   \right)\leq 2|\cX||\cY||\cZ| e^{-n K} +\epsilon.
\end{equation*}
Here $K>0$ and $K$ does not depend on $n$, $P_{X,Y}$, $P_{\vec{Z}}$ or $(\vec{x},\vec{y})$ but does depend on $\dnot$, $g$ and $P_{Z|Y}$. Further, the $\delta$ function does not depend on $(\bx,\by)$, $P_{X,Y}$ or $P_{\bZ}$.
\end{lemma}
We now use the above lemma to prove the following claim.
\begin{claim}\label{claim:x:2:u:typ}
 There exists $\dfourd>0$, where $\dfourd\rightarrow 0$ as
$\delta\rightarrow 0$, such that except for a small
probability, $(\bX,\bJ,\bZ,\bY,\bU)$
is jointly $\dfour$-typical w.r.t. $P_XT_{\bJ|\bX}W_{Y,Z|X,J}P_{U|Y}$.
\end{claim}
\begin{IEEEproof}
Let us assume that $A_{X,J,Y,Z}$ is true.
Now we use the refined Markov lemma (Lemma~\ref{lem:ref:markov:lemma}) on the Markov chain $(X,J,Z) \rightarrow
Y \rightarrow U$.
Then, by Claims~\ref{claim:y:u:tpy} and~\ref{claim:unif:dist}, $\bU$ is chosen such that both
conditions (a) and (b) in Lemma~\ref{lem:ref:markov:lemma} are satisfied.
Thus, the claim follows.
\end{IEEEproof}
We define this ``good'' event as $A_{X,J,Y,Z,U}:=\{(\bX,\bJ,\bZ,\bY,\bU)
\in \cT^n_{\dfour}(P_XT_{\bJ|\bX}W_{YZ|XJ}P_{U|Y}\}$.
\begin{claim}
There exists $\gamma(\delta)>0$, where $\gamma(\delta)\rightarrow 0$ as $\delta\rightarrow 0$, such that except for an exponentially small probability, $\bU\in \cL_{\gamma(\delta)}(M,\bZ)$. 
\end{claim}
\begin{IEEEproof}
Consider the event $A_{X,J,Y,Z,U}$. Under this event, $(\bU,\bZ)$ are
$\gammad$-typical w.r.t. $P_{U,Z}=[P_X T_{\bJ|\bX} W_{Y,Z|X,J}
P_{U|Y}]_{U,Z}$, where $\gammad=|\cX||\cJ| |\cY| \dfour $. 
Thus, the claim follows from Claim~\ref{claim:x:2:u:typ}.
\end{IEEEproof}
%
This completes the proof of the first part of the lemma. The proof of the second part directly follows from the following claim.
\begin{claim}
There exists $f_2(\delta,\epsilon)>0$, where $f_2(\delta,\epsilon)\rightarrow 0$ as $\delta,\epsilon\rightarrow 0$, such that
\begin{IEEEeqnarray}{rCl}
\mathbb{P}\left(\vec{U}_{M,L'}\in \mathcal{L}_{\gamma(\delta)}(M,\vec{Z}), \text{ for some } L'\neq L \right)\leq 2^{-n f_2(\delta,\epsilon)}.
\end{IEEEeqnarray}
\end{claim}
\begin{IEEEproof}
Note that the codewords $\{\bU_{M,L'}\}_{L'\neq L}$ are independently generated, and hence, $\{\bU_{M,L'}\}_{L'\neq L}$ and $\bZ$ are independent. Consider a fixed conditional type $T_{J|X}\in \sQ(T_{\bY})$, and let the resulting distribution $P_{U,Z}=[P_X T_{J|X} W_{Y,Z|X,J} P_{U|Y}]_{U,Z}$. Then,
\begin{align*}
\mathbb{P}\big(\exists l'\neq L:(\vec{U}_{M,l'},\bZ)\in \cT^n_{\gamma(\delta)} (P_{U,Z})) \leq  2^{-n\tdftwo}
\end{align*}
for some $\td{\ftwo} \rightarrow 0$ as $\delta, \epsilon \rightarrow 0$. This follows from the packing lemma~\cite[Lemma~3.1]{elgamal-kim}. 
By taking the union bound over all conditional types $T_{J|X}\in\sQ(T_{\bY})$ (the number of such types is at most polynomial in $n$), we get
\begin{IEEEeqnarray*}{rCl}
\mathbb{P}\big(\exists l'\neq L: (\vec{U}_{M,l'},\bZ)\in \cT^n_{\gamma(\delta)} (P_{U,Z}) \text{  for some } T_{J|X}\in \sQ(T_{\bY})) &\leq& (n+1)^{|\cU||\cZ|} 2^{-n\tdftwo}\\
&\leq & 2^{-n\ftwo}.
\end{IEEEeqnarray*}
\end{IEEEproof}
This completes the proof of the lemma.

\section{}\label{app:sec:converse}
Here we prove~\eqref{eq:wyn}. We first state the following useful lemma.
\begin{lemma}\label{lem:f:d:props}
For a fixed $Q_{J|X}\in\sQ$, $F(D,Q_{J|X})$ is a non-decreasing, convex and continuous function of $D$.
\end{lemma}
\begin{IEEEproof}
The proof of Lemma~\ref{lem:f:d:props} can be given using similar arguments as
in the proof of the same statement about the rate-distortion function
for the Wyner-Ziv problem~\cite{wyner-ziv-it1976}. We provide the proof 
below for completeness.

To prove that $F(D,Q_{J|X})$ is a non-increasing function of $D$, note that
the minimization in the definition of $F(D,Q_{J|X})$ is over
the set $\mathcal{S}=\{(P_{U|Y},\tilde{x}(.,.)):\mathbb{E}[d(X,\wtd{X})]\leq
D$\}. So, for $D'_2> D'_1$, the corresponding domains of minimization satisfy
$\mathcal{S}_2 \supseteq \mathcal{S}_1$. Thus $F(D'_2,Q_{J|X})\leq
F(D'_1,Q_{J|X})$.

To prove the convexity of $F(D,Q_{J|X})$ as a function of $D$,
note that the minimization over $P_{U|Y}$ and
$\td{x}(.,.)$ can be rewritten as a minimization over only $P_{U|Y}$ in a
manner similar to~\cite{willems-report}. Then, the alphabet of the auxiliary
random variable $U$ is the set of `Shannon strategies'.
To establish the convexity, we will first show that
for given $Q_{J|X}\in\sQ$ and for fixed $P_X$ and $W_{Y,Z|X,J}$, $I(U;Y|Z)$ is
convex in $P_{U|Y}$.
Toward this, consider the joint distribution $P_{X,Y,Z,U}=[P_{X} Q_{J|X}
W_{Y,Z|X,J} P_{U|Y}]_{X,Y,Z,U}$. For fixed $P_{Z}$, we know that $I(U;Y|Z)$ is a convex
function of $P_{U,Y|Z}$~\cite{cover-thomas}. Now $P_{U,Y|Z}=P_{Y|Z}
P_{U|Y,Z}=P_{Y|Z} P_{U|Y}$, where the last equality follows from the Markov
chain $(X,J,Z)\rightarrow Y\rightarrow U$. As $P_X$, $Q_{J|X}$ and
$W_{Y,Z|X,J}$ are fixed, it follows that $P_{Y,Z}$ is fixed, and hence,
$I(U;Y|Z)$ is a convex function of $P_{U|Y}$.

Now consider two distortion values $D'_1,D'_2$, such that 
$P^{(1)}_{U|Y}$ and $P^{(2)}_{U|Y}$ achieve the values
$F(D'_1,Q_{J|X})$ and $F(D'_2,Q_{J|X})$ respectively.
Let us define the convex combinations $D_\lambda = 
\lambda D'_1+(1-\lambda) D'_2$ and $P^{(\lambda)}_{U|Y}
= \lambda P^{(1)}_{U|Y} + (1-\lambda)P^{(2)}_{U|Y}$. Note that
the other factors of the joint distribution 
$P_{X,J,Y,Z,U}=P_X Q_{J|X} W_{Y,Z|X,J} P_{U|Y}$ are fixed here.
So the average distortion is linear in $P_{U|Y}$, and thus,
$P^{(\lambda)}_{U|Y}$ is a feasible distribution for $D_\lambda$. Thus
\begin{IEEEeqnarray*}{rCl}
F(D_\lambda,Q_{J|X}) &\stackrel{}{\leq}& I_{P^{(\lambda)}}(U;Y|Z)\\
&\stackrel{(a)}{\leq}&  \lambda I_{P^{(1)}  }(U;Y|Z)+(1-\lambda)I_{P^{(2)}} (U;Y|Z)\\
&\stackrel{}{=}&  \lambda F(D'_1,Q_{J|X})+ (1-\lambda) F(D'_2,Q_{J|X}).\yesnumber\label{eq:con}
\end{IEEEeqnarray*}
Here $(a)$ follows from the convexity of $I(U;Y|Z)$ w.r.t. $P_{U|Y}$.
This proves the convexity of $F(D,Q_{J|X})$ w.r.t. $D$.

Finally, the continuity of $F(D,Q_{J|X})$ follows from its convexity~\cite{rockafeller-book1972}.
\end{IEEEproof}

In the following, for the given code and the i.i.d. jamming distribution
$Q_{J|X}$, we denote $D_i :=E[d(X_i,\widetilde{X}_i)]$.
Now we have
\begin{IEEEeqnarray*}{rCl}
nR
&\stackrel{}{\geq}& H(M|\vec{Z},\Theta)\\
&\stackrel{}{=}& I(\vec{Y};M|\vec{Z},\Theta)\\
&\stackrel{}{= }& \sum_{i=1}^n I(Y_i;M|\vec{Y}^{i-1},\vec{Z},\Theta)\\
&\stackrel{(a)}{= }& \sum_{i=1}^n I(Y_i;M,\vec{Y}^{i-1},\vec{Z}^{i-1},\bZ_{i+1}^n,\Theta|Z_i)\\
&\stackrel{(b)}{= }& \sum_{i=1}^n I(Y_i;U_i|Z_i)\\
&\stackrel{(c)}{\geq}& \sum_{i=1}^n \min_{P_{U_i|Y_i},P_{\wtd{X}_i|U_i,Z_i} } I(U_i;Y_i|Z_i)\\
&\stackrel{(d)}{=}& n\left(\sum_{i=1}^n \frac{1}{n}F(D_i, Q_{J|X})\right)\\
&\stackrel{(e)}{\geq}& nF\left( \sum_{i=1}^n \frac{D_i}{n}, Q_{J|X}\right)\\
&\stackrel{(f)}{\geq}& nF({D^{(n)}},Q_{J|X}).\label{eq:r:fd}
\end{IEEEeqnarray*}
Under the memoryless jamming strategy of the adversary given in~\eqref{eq:mem:jam}, $Y_i$ and 
$(\vec{Z}^{i-1},\bZ_{i+1}^n,\vec{Y}^{i-1},\Theta)$ are independent, conditioned
on $Z_i$. This gives us $(a)$. By defining
$U_i=(M,\vec{Y}^{i-1},\vec{Z}^{i-1},\bZ_{i+1}^n,\Theta)$, we have $(b)$.  
For $(c)$, note first that from~\eqref{eq:mem:jam}, it follows that
$U_i\leftrightarrow Y_i\leftrightarrow Z_i,~\forall i$. The minimization in (c)
can hence be taken over pairs $(P_{U_i|Y_i},P_{\wtd{X}_i|U_i,Z_i})$ such that
$E[d(X_i,\widetilde{X}_i)]\leq D_i$ under $Q_{J|X}$.  We get $(d)$ from the
fact that the distortion constraint in the minimization being linear in
$P_{\tilde{X}_i|U_i,Z_i}$ allows us to replace it with a function
$\tilde{x}(.,.)$ of $u,z$ as in the definition of $F(D,Q_{J|X})$
in~\eqref{eq:ef}.  Finally, $(e)$ and $(f)$ follow respectively from the
convexity and non-decreasing nature of $F(D,Q_{J|X})$ (from
Lemma~\ref{lem:f:d:props}) where we note that $(1/n) \sum_i D_i\leq D^{(n)}$.
This establishes~\eqref{eq:wyn}.

%
\bibliographystyle{IEEEtran}

\bibliography{IEEEabrv,References}

\begin{thebibliography}{10}
\providecommand{\url}[1]{#1}
\csname url@samestyle\endcsname
\providecommand{\newblock}{\relax}
\providecommand{\bibinfo}[2]{#2}
\providecommand{\BIBentrySTDinterwordspacing}{\spaceskip=0pt\relax}
\providecommand{\BIBentryALTinterwordstretchfactor}{4}
\providecommand{\BIBentryALTinterwordspacing}{\spaceskip=\fontdimen2\font plus
\BIBentryALTinterwordstretchfactor\fontdimen3\font minus
  \fontdimen4\font\relax}
\providecommand{\BIBforeignlanguage}[2]{{%
\expandafter\ifx\csname l@#1\endcsname\relax
\typeout{** WARNING: IEEEtran.bst: No hyphenation pattern has been}%
\typeout{** loaded for the language `#1'. Using the pattern for}%
\typeout{** the default language instead.}%
\else
\language=\csname l@#1\endcsname
\fi
#2}}
\providecommand{\BIBdecl}{\relax}
\BIBdecl

\bibitem{lapidoth-narayan-it1998}
A.~Lapidoth and P.~Narayan, ``Reliable communication under channel
  uncertainty,'' \emph{{IEEE} Trans. Inform. Theory}, vol.~44, pp. 2148--2177,
  1998.

\bibitem{dobrushin-tsybakov-it1962}
R.~Dobrushin and B.~Tsybakov, ``Information transmission with additional
  noise,'' \emph{IRE Trans. Inform. Theory}, vol.~8, pp. 293--304, September
  1962.

\bibitem{wyner-ziv-it1976}
A.~D. Wyner and J.~Ziv, ``The rate-distortion function for source coding with
  side information at the decoder,'' \emph{{IEEE} Trans. Inform. Theory},
  vol.~22, pp. 1--10, January 1976.

\bibitem{shannon-ire1959}
C.~E. Shannon, ``Coding theorems for a discrete source with a fidelity
  criterion,'' \emph{IRE Nat. Conv. Rec.}, vol.~7, pp. 142--163, 1959.

\bibitem{kieffer-it1993}
J.~C. Kieffer, ``A survey of the theory of source coding with a fidelity
  criterion,'' \emph{{IEEE} Trans. Inform. Theory}, vol.~39, pp. 1473--1490,
  September 1993.

\bibitem{berger-gibson-it1998}
T.~Berger and J.~D. Gibson, ``Lossy source coding,'' \emph{{IEEE} Trans.
  Inform. Theory}, vol.~44, pp. 2693--2723, October 1998.

\bibitem{dembo-weissman-it1993}
A.~Dembo and T.~Weissman, ``The minimax distortion redundancy in noisy source
  coding,'' \emph{{IEEE} Trans. Inform. Theory}, vol.~49, pp. 3020--3030,
  November 2003.

\bibitem{neuhoff-etal-it1975}
D.~L. Neuhoff, R.~M. Gray, and L.~D. Davisson, ``Fixed rate universal block
  source coding with a fidelity criterion,'' \emph{{IEEE} Trans. Inform.
  Theory}, vol.~21, pp. 511--523, September 1975.

\bibitem{sakrison-inc1969}
D.~Sakrison, ``The rate distortion function for a class of sources,''
  \emph{Information and Control}, vol.~15, pp. 165--195, 1969.

\bibitem{watanabe-etal-it2014}
S.~Watanabe and S.~Kuzuoka, ``Universal {W}yner-{Z}iv coding for distortion
  constrained general side information,'' \emph{{IEEE} Trans. Inform. Theory},
  vol.~60, pp. 7568--7583, December 2014.

\bibitem{csiszar-korner-book2011}
I.~Csisz\'{a}r and J.~K{\"o}rner, \emph{Information theory: coding theorems for
  discrete memoryless systems}.\hskip 1em plus 0.5em minus 0.4em\relax
  Cambridge University Press, 2011.

\bibitem{feder-merhav-isit1995}
M.~Feder and N.~Merhav, ``Universal coding for arbitrarily varying sources,''
  in \emph{Proc. {IEEE} Int. Symp. Information Theory}, Whistler, Canada,
  September 1995.

\bibitem{berger-it1971}
T.~Berger, ``The source coding game,'' \emph{{IEEE} Trans. Inform. Theory},
  vol.~17, pp. 71--76, January 1971.

\bibitem{palaiyanur-it2011}
H.~Palaiyanur, C.~Chang, and A.~Sahai, ``The source coding game with a cheating
  switcher,'' \emph{{IEEE} Trans. Inform. Theory}, vol.~57, pp. 4545--4560,
  July 2011.

\bibitem{willems-report}
F.~M.~J. Willems, ``Computation of the {W}yner--{Z}iv rate-distortion
  function,'' Technische Hogeschool Eindhoven, Tech. Rep., 1983.

\bibitem{sion-pjm1958}
M.~Sion, ``On general minimax theorems,'' \emph{Pacific J. Math.}, vol.~8, pp.
  171--176, 1958.

\bibitem{bdp-arxiv2016}
\BIBentryALTinterwordspacing
A.~J. Budkuley, B.~K. Dey, and V.~M. Prabhakaran, ``Communication in the
  presence of a state-aware adversary,'' \emph{Arxiv}, 2016. [Online].
  Available: \url{arxiv.org/pdf/1509.08299}
\BIBentrySTDinterwordspacing

\bibitem{elgamal-kim}
A.~E. Gamal and Y.-H. Kim, \emph{Network Information Theory}.\hskip 1em plus
  0.5em minus 0.4em\relax Cambridge University Press, 2011.

\bibitem{hoeffding-1963}
W.~Hoeffding, ``Probability inequalities for sums of bounded random
  variables,'' \emph{Journal of the American Statistical Association}, vol.~58,
  pp. 13--30, 1963.

\bibitem{cover-thomas}
T.~Cover and J.~Thomas, \emph{Elements of Information Theory}.\hskip 1em plus
  0.5em minus 0.4em\relax Wiley, New York, 1991.

\bibitem{rockafeller-book1972}
R.~T. Rockafeller, \emph{Convex Analysis}.\hskip 1em plus 0.5em minus
  0.4em\relax Princeton University Press, 1972.

\end{thebibliography}

\end{document}